\documentclass[10pt]{article}
\usepackage{graphicx}
\usepackage{amsmath}
\usepackage{amssymb}
\usepackage{caption2}
\begin{document}
\begin{center}
{\large {\bf \sc{  Analysis of the doubly-charmed tetraquark molecular states with the QCD sum rules }}} \\[2mm]
Qi Xin, Zhi-Gang  Wang \footnote{E-mail: zgwang@aliyun.com.  }     \\
 Department of Physics, North China Electric Power University, Baoding 071003, P. R. China
\end{center}

\begin{abstract}
In the present work, we investigate the scalar, axialvector and tensor doubly-charmed  tetraquark molecular states  without strange, with strange and with doubly-strange via the QCD sum rules, and try to make assignment of the $T^+_{cc}$ from the LHCb collaboration in the scenario of  molecular states.  The predictions favor assigning the $T^+_{cc}$ to be the lighter $DD^{*}$  molecular state with the spin-parity $J^P=1^+$ and isospin $I=0$, while the heavier $DD^{*}$  molecular state with the spin-parity $J^P=1^+$ and isospin $I=1$ still escapes  experimental detections, the observation of the heavier $DD^{*}$  molecular state would shed light on the nature of the $T_{cc}^+$. All the predicted doubly-charmed tetraquark  molecular states can be confronted   to the experimental data in the future.
 \end{abstract}

PACS number: 12.39.Mk, 12.38.Lg

Key words: Tetraquark molecular  state, QCD sum rules

\section{Introduction}

In recent years,  a series of   charmonium-like (and bottomonium-like) states were observed  by the  international high energy experiments \cite{PDG},   those states lie nearby  the thresholds of two charmed (or bottom) mesons and have hidden-charm (or hidden-bottom), and are potential excellent candidates for the tetraquark states or  tetraquark molecular states. Up to now, those exotic $X$, $Y$, $Z$ states have  posed a big challenge and a big opportunity for the hadron spectroscopy, and provide us with an active research field as an extension of the conventional quark model.

In 2007, the  Belle collaboration   observed the $Z_c(4430)$  in the $\pi^{\pm} \psi^\prime$  mass spectrum  in the $B\to K \pi^{\pm} \psi^{\prime}$ decays \cite{Belle-2007}.
In 2014, the LHCb collaboration   provided the first independent confirmation of the existence of the $Z_c^-(4430)$
and established its spin-parity  $J^P=1^+$ by analyzing  the $B^0\to\psi'\pi^-K^+$ decays \cite{LHCb-1404}.
In 2013, the BESIII collaboration (also the Belle collaboration)  observed the  $Z_c(3900)$ in the $\pi^\pm J/\psi$ mass spectrum in  the process  $e^+e^- \to \pi^+\pi^-J/\psi$  \cite{BES3900} (\cite{Belle3900}). The $Z_c(4430)$ and $Z_c(3900)$ have non-zero electric charge and have the valence quarks $c\bar{c}u\bar{d}$
or $c\bar{c}d\bar{u}$, and they are very good candidates for the hidden-charm tetraquark (molecular) states.

In 2015,  the  LHCb collaboration  observed  the pentaquark candidates $P_c(4380)$ and $P_c(4450)$ in the $J/\psi p$ mass spectrum  in the $\Lambda_b^0\to J/\psi K^- p$ decays   \cite{LHCb-4380}. In 2019, the LHCb collaboration  observed the pentaquark candidate $P_c(4312)$  and confirmed the  pentaquark structure $P_c(4450)$, and proved  that it consists  of two narrow overlapping peaks $P_c(4440)$ and $P_c(4457)$  \cite{LHCb-Pc4312}. The $P_c(4312)$, $P_c(4440)$ and $P_c(4457)$ have positive electric charge and have the valence quarks $c\bar{c}uud$, and they are very good candidates for the hidden-charm pentaquark (molecular) states.

In 2017, the LHCb collaboration observed the doubly-charmed baryon  $\Xi_{cc}^{++}$ in the $\Lambda_c^+ K^- \pi^+\pi^+$ mass spectrum  \cite{LHCb-Xicc}.
The observation of the $\Xi_{cc}^{++}$ provides us with  the crucial experimental input on the strong correlation between the two charm quarks, which is of great importance on the spectroscopy of the doubly-charmed  baryon states, tetraquark states and pentaquark states to gain a deeper insight
into the mechanism  of low energy QCD, and stimulates  many interesting works, for example, the doubly-heavy pentaquark states \cite{WZG-QQ-penta-EPJC}.

In the same road, very recently, the LHCb collaboration reported an important observation of the doubly-charmed tetraquark candidate $T_{cc}^+$ in the $D^0D^0\pi^+$  mass spectrum at the European Physical Society conference on "high energy physics 2021" \cite{Tcc:talk}. Subsequently, the LHCb collaboration formally announced the  observation of the exotic state $T_{cc}^+$   just below  the $D^0D^{*+}$ threshold using a data set corresponding to
  an integrated luminosity of $9\rm{fb}^{-1}$ acquired  with the LHCb detector in   proton-proton collisions  at center-of-mass
  energies of 7, 8 and 13 $\rm{TeV}$ \cite{LHCb-Tcc,LHCb-Tcc-detail}.  The Breit-Wigner mass and width are $\delta M_{BW} = -273\pm 61\pm 5^{+11}_{-14}~\text{KeV}$ below the $D^0D^{*+}$ threshold and $\Gamma_{BW} = 410\pm 165\pm 43^{+18}_{-38}~\text{KeV}$  \cite{Tcc:talk,LHCb-Tcc,LHCb-Tcc-detail}. The exotic state $T^+_{cc}$ is consistent with  the ground state isoscalar  tetraquark state with  a valence quark content of  $cc\bar{u}\bar{d}$   and spin-parity  $J^{P}=1^+$, and  exploring the $DD$ mass spectrum  disfavors
 interpreting  the  $T_{cc}^+$ as the isovector state.
 The observation of the  $T_{cc}^+$ is a great breakthrough  beyond the $\Xi_{cc}^{++}$ for the hadron physics, and it is the firstly observed doubly-charmed tetraquark candidate  with the typical quark configuration $cc\bar{u}\bar{d}$, it is a very good candidate for the tetraquark (molecular) state having doubly-charm.

Before and after the observation of the doubly-charmed tetraquark candidate $T_{cc}^+$, there have been several works on the doubly-charmed tetraquark (molecular) states using  the QCD sum rules \cite{QQ-QCDSR-Narison, QQ-QCDSR-WZG-CTP,QQ-QCDSR-Chen-Zhu,QQ-QCDSR-WZG-APPB,QQ-QCDSR-WZG-EPJC,QQ-QCDSR-Azizi-NPB,QQ-QCDSR-Azizi},
the nonrelativistic quark model \cite{QQ-NRQM-Semay,QQ-NRQM-Vijande,QQ-NRQM-Lee,QQ-NRQM-Yang,QQ-NRQM-Karliner,QQ-NRQM-Esposito} (\cite{QQ-NRQM-mole-Janc}), (the coupled-channel analysis \cite{QQ-CCA-mole-Zhu,QQ-CCA-mole-HeJ,QQ-CCA-mole-Zhu-CPL,QQ-CCA-mole-Meng,QQ-CCA-mole-RChen,QQ-CCA-mole-Liang,QQ-CCA-mole-GuoFK}), the heavy quark symmetry \cite{QQ-HQS-Eichten,QQ-HQS-Cheng} (\cite{QQ-HQS-mole-Liu}), the lattice QCD \cite{QQ-Latt-Junnarkar}, (the effective Lagrangian approach \cite{QQ-ELA-Geng}), etc.
The predicted masses vary from  about $250 \,\rm{MeV}$ below to $250\,\rm{MeV}$ above the mass of the measured value of the $M_{T_{cc}^+}$.
The closeness to the $ DD^*$ threshold makes one think immediately about the possibility of assigning  it to be a $DD^*$ molecular state.

In the past years, the QCD sum rules have  become a powerful theoretical approach in exploring  the masses and decay widths of the $X$, $Y$ and $Z$ states to diagnose their natures, irrespective of the  hidden-charm (or hidden-bottom) tetraquark states  or tetraquark  molecular states. In Refs.\cite{QQ-QCDSR-Narison, QQ-QCDSR-WZG-CTP,QQ-QCDSR-Chen-Zhu,QQ-QCDSR-WZG-APPB,QQ-QCDSR-WZG-EPJC,QQ-QCDSR-Azizi-NPB,QQ-QCDSR-Azizi}, the diquark-antidiquark type doubly-heavy tetraquark
states are investigated by  the QCD sum rules.  In Refs.\cite{QQ-QCDSR-WZG-APPB,QQ-QCDSR-WZG-EPJC}, we investigate  the  scalar, axialvector, vector, tensor
doubly-charmed tetraquark states with QCD sum rules  systematically  by carrying out the operator product expansion up to the vacuum condensates of dimension 10 in a consistent way and taking the energy scale formula to determine the suitable energy scales of the QCD spectral densities. The predicted masses for  the axialvector-diquark-scalar-antidiquark type and axialvector-diquark-axialvector-antidiquark type  axialvector doubly-charmed tetraquark states $cc\bar{u}\bar{d}$ are  $3.90\pm0.09\,\rm{GeV}$, which is in excellent agreement with the experimental value from the LHCb collaboration \cite{Tcc:talk,LHCb-Tcc,LHCb-Tcc-detail}.

If we perform Fierz rearrangements for the axialvector-diquark-scalar-antidiquark type  four-quark axialvector  currents $J_\mu(x)$ both in the color and Dirac spinor spaces \cite{QQ-QCDSR-WZG-APPB}, we can obtain  special superpositions of the color-singlet-color-singlet type currents,
\begin{eqnarray}\label{Fierz}
J_\mu(x)&=&\varepsilon^{ijk}\varepsilon^{imn} \, Q^{T}_j(x)C\gamma_\mu Q_k(x) \,\bar{u}_m(x)\gamma_5C \bar{d}^T_n(x) \, , \nonumber\\
&=&\frac{i}{2}\left[\bar{u}i\gamma_5Q \bar{d}\gamma_\mu Q -\bar{d}i\gamma_5 Q\bar{u}\gamma_\mu Q \right]+\frac{1}{2}\left[\bar{u}Q \bar{d}\gamma_\mu\gamma_5 Q -\bar{d} Q\bar{u}\gamma_\mu \gamma_5Q \right]\nonumber\\
&&-\frac{i}{2}\left[\bar{u}\sigma_{\mu\nu}\gamma_5Q \bar{d}\gamma^\nu Q -\bar{d}\sigma_{\mu\nu}\gamma_5 Q\bar{u}\gamma^\nu Q \right]+\frac{i}{2}\left[\bar{u}\sigma_{\mu\nu}Q \bar{d}\gamma^\nu \gamma_5Q -\bar{d}\sigma_{\mu\nu} Q\bar{u}\gamma^\nu\gamma_5 Q \right]\, , \nonumber\\
&=&\frac{i}{2}J_\mu^1(x)+\frac{1}{2}J_\mu^2(x)-\frac{i}{2}J_\mu^3(x)+\frac{i}{2}J_\mu^4(x)\, ,
\end{eqnarray}
where $Q=c,b$, the  $i$, $j$, $k$,  $m$, $n$ are color indexes, the $C$ is the charge conjugation matrix. The currents $J_\mu^1(x)$, $J_\mu^2(x)$, $J_\mu^3(x)$ and $J_\mu^4(x)$ couple potentially to the color-singlet-color-singlet type tetraquark states or two-meson scattering states. The decays to the two-meson states can take place through the Okubo-Zweig-Iizuka super-allowed fall-apart mechanism if they are allowed in the phase-space.

In fact, there exist spatial separations between the diquark and antidiquark pair, the currents $J_\mu(x)$ should be modified to $J_\mu(x,\epsilon)$,
\begin{eqnarray}
J_\mu(x,\epsilon)&=&\varepsilon^{ijk}\varepsilon^{imn} \, Q^{T}_j(x)C\gamma_\mu Q_k(x) \,\bar{u}_m(x+\epsilon)\gamma_5C \bar{d}^T_n(x+\epsilon) \, ,
\end{eqnarray}
where the four-vector $\epsilon^\alpha=(0,\vec{\epsilon})$. The repulsive barrier or spatial distance between the diquark and antidiquark pair frustrates the Fierz rearrangements or recombinations \cite{Wilczek-diquark,Polosa-diquark,Maiani-1903,Brodsky-PRL,WZG-nonlocal}, though in practical calculations we usually take the local limit $\epsilon \to 0$, we should not take it for granted that the Fierz rearrangements are feasible \cite{WZG-nonlocal}, so we cannot obtain the conclusion that the $T_{cc}^+$ has the $D^{*}D -DD^{*}$ Fock component according to the properties of the currents $\bar{q}i\gamma_5c\otimes\bar{q}\gamma_\mu c$ in Eq.\eqref{Fierz}. At most,  we can only obtain the conclusion tentatively that the $T_{cc}^+$ has a diquark-antidiquark type tetraquark Fock component with the spin-parity $J^P=1^+$ and isospin $I=0$ \cite{QQ-QCDSR-WZG-APPB}. It is interesting to explore whether or not there exists a color-singlet-color-singlet type tetraquark Fock component for the $T_{cc}^+$ state, or in the other words, whether or not there exists a color-singlet-color-singlet type tetraquark state which has a mass about $3875\,\rm{MeV}$ based on the QCD sum rules.

In the present work,  we extend our previous works \cite{QQ-QCDSR-WZG-APPB,QQ-QCDSR-WZG-EPJC} to explore  the color-singlet-color-singlet type scalar, axialvector and tensor  doubly-charmed tetraquark  states  via the QCD sum rules by accomplishing   the operator product expansion up to the vacuum condensates of dimension $10$ and taking  account of the $SU(3)$ breaking effects in a consistent way, and make possible assignment of the $T_{cc}^+$ in the scenario of the hadronic  molecules (or color-singlet-color-singlet type tetraquark states) based on the QCD sum rules. Furthermore, we obtain reliable predictions for  the doubly-charmed scalar, axialvector and tensor tetraquark molecular states without strange, with strange and with doubly-strange, which can be confronted to the experimental data in the future.

The article is arranged as follows:  we acquire the QCD sum rules for the masses and pole residues  of  the  doubly-charmed tetraquark molecular states in section 2; in section 3, we   present the numerical results and discussions; section 4 is reserved for our conclusion.

\section{QCD sum rules for  the  doubly-charmed  tetraquark molecular states}

Firstly, let us  write down  the two-point correlation functions $\Pi(p)$, $\Pi_{\mu\nu}(p)$ and $\Pi_{\mu\nu\alpha\beta}(p)$  in the QCD sum rules,
\begin{eqnarray}\label{CF-Pi}
\Pi(p)&=&i\int d^4x e^{ip \cdot x} \langle0|T\Big\{J(x)J^{\dagger}(0)\Big\}|0\rangle \, ,\nonumber\\
\Pi_{\mu\nu}(p)&=&i\int d^4x e^{ip \cdot x} \langle0|T\Big\{J_\mu(x)J_{\nu}^{\dagger}(0)\Big\}|0\rangle \, ,\nonumber\\
\Pi_{\mu\nu\alpha\beta}(p)&=&i\int d^4x e^{ip \cdot x} \langle0|T\Big\{J_{\mu\nu}(x)J_{\alpha\beta}^{\dagger}(0)\Big\}|0\rangle \, ,
\end{eqnarray}
where the currents
\begin{eqnarray}
J(x)&=&J_{{D}^*{D}^*}(x)\, , J_{D^*{D}_s^*}(x)\, , J_{{D}_s^*{D}_s^*}(x)\, ,
\end{eqnarray}
\begin{eqnarray}
J_{\mu}(x)&=&J_{D{D}^*,L,\mu}(x)\, , \, J_{D{D}^*,H,\mu}(x)\, ,
 J_{D{D}_s^*,L,\mu}(x)\, , \, J_{D{D}_s^*,H,\mu}(x)\, , \,  J_{D_s{D}_s^*,\mu}(x)\, , \nonumber\\
 && J_{D_1D_0^*,L,\mu}(x)\, , \,  J_{D_1D_0^*,H,\mu}(x)\, , \,  J_{D_{s1}D_{0}^*,L,\mu}(x)\, , \, J_{D_{s1}D_{0}^*,H,\mu}(x)\, , \,  J_{D_{s1}D_{s0}^*,\mu}(x)\, ,
 \end{eqnarray}
\begin{eqnarray}
 J_{\mu\nu}(x)&=&J_{{D}^*{D}^*,L,\mu\nu}(x)\, , \, J_{{D}^*{D}^*,H,\mu\nu}(x)\, , \,   J_{{D}^*{D}_s^*,L,\mu\nu}(x)\, , \, J_{{D}^*{D}_s^*,H,\mu\nu}(x)\, , \,J_{{D}_s^*{D}_s^*,L,\mu\nu}(x)\, ,  \nonumber\\
 &&  J_{{D}_s^*{D}_s^*,H,\mu\nu}(x)\, ,
\end{eqnarray}
\begin{eqnarray}
J_{{D}^*{D}^*}(x)&=&\bar{u}(x)\gamma_\mu c(x)\, \bar{d}(x)\gamma^\mu c(x) \, ,\nonumber\\
J_{{D}^*{D}_s^*}(x)&=&\bar{q}(x)\gamma_\mu c(x)\, \bar{s}(x)\gamma^\mu c(x) \, ,\nonumber\\
J_{{D}^*_s{D}_s^*}(x)&=&\bar{s}(x)\gamma_\mu c(x)\, \bar{s}(x)\gamma^\mu c(x) \, ,
\end{eqnarray}
\begin{eqnarray}
J_{D{D}^*,L,\mu}(x)&=&\frac{1}{\sqrt{2}}\Big[\bar{u}(x)i\gamma_5c(x)\, \bar{d}(x)\gamma_\mu c(x) -\bar{u}(x)\gamma_\mu c(x)\,\bar{d}(x)i\gamma_5 c(x)\,  \Big] \, ,\nonumber\\
J_{D{D}^*,H,\mu}(x)&=&\frac{1}{\sqrt{2}}\Big[\bar{u}(x)i\gamma_5c(x)\, \bar{d}(x)\gamma_\mu c(x) +\bar{u}(x)\gamma_\mu c(x)\,\bar{d}(x)i\gamma_5 c(x)  \Big] \, ,\nonumber\\
J_{D{D}_s^*,L,\mu}(x)&=&\frac{1}{\sqrt{2}}\Big[\bar{q}(x)i\gamma_5c(x)\, \bar{s}(x)\gamma_\mu c(x)-\bar{q}(x)\gamma_\mu c(x)\, \bar{s}(x)i\gamma_5 c(x) \Big] \, ,\nonumber\\
J_{D{D}_s^*,H,\mu}(x)&=&\frac{1}{\sqrt{2}}\Big[\bar{q}(x)i\gamma_5c(x)\, \bar{s}(x)\gamma_\mu c(x)+\bar{q}(x)\gamma_\mu c(x)\, \bar{s}(x)i\gamma_5 c(x) \Big] \, ,  \nonumber\\
J_{D_s{D}_s^*,\mu}(x)&=&\bar{s}(x)i\gamma_5c(x)\, \bar{s}(x)\gamma_\mu c(x) \, ,
\end{eqnarray}
\begin{eqnarray}
J_{D_1D_0^*,L,\mu}(x)&=&\frac{1}{\sqrt{2}}\Big[\bar{u}(x)c(x)\, \bar{d}(x)\gamma_\mu \gamma_5c(x) +\bar{u}(x)\gamma_\mu \gamma_5  c(x)\, \bar{d}(x)c(x) \Big] \, ,\nonumber\\
J_{D_1D_0^*,H,\mu}(x)&=&\frac{1}{\sqrt{2}}\Big[\bar{u}(x)c(x)\, \bar{d}(x)\gamma_\mu \gamma_5 c(x) -\bar{u}(x)\gamma_\mu \gamma_5 c(x)\, \bar{d}(x) c(x) \Big] \, ,\nonumber\\
J_{D_{s1}D_{0}^*,L,\mu}(x)&=&\frac{1}{\sqrt{2}}\Big[\bar{q}(x)c(x)\, \bar{s}(x)\gamma_\mu \gamma_5 c(x)+\bar{q}(x)\gamma_\mu \gamma_5 c(x)\, \bar{s}(x) c(x) \Big] \, ,\nonumber\\
J_{D_{s1}D_{0}^*,H,\mu}(x)&=&\frac{1}{\sqrt{2}}\Big[\bar{q}(x)c(x)\, \bar{s}(x)\gamma_\mu \gamma_5 c(x)-\bar{q}(x)\gamma_\mu \gamma_5 c(x)\, \bar{s}(x) c(x) \Big] \, , \nonumber\\
J_{D_{s1}D_{s0}^*,\mu}(x)&=&\bar{s}(x)c(x)\, \bar{s}(x)\gamma_\mu \gamma_5 c(x) \, ,
\end{eqnarray}
\begin{eqnarray}
J_{{D}^*{D}^*,L,\mu\nu}(x)&=&\frac{1}{\sqrt{2}}\Big[\bar{u}(x)\gamma_\mu c(x)\, \bar{d}(x)\gamma_\nu c(x) -\bar{u}(x)\gamma_\nu c(x)\,\bar{d}(x)\gamma_\mu c(x)  \Big] \, ,\nonumber\\
J_{{D}^*{D}^*,H,\mu\nu}(x)&=&\frac{1}{\sqrt{2}}\Big[\bar{u}(x)\gamma_\mu c(x)\, \bar{d}(x)\gamma_\nu c(x) +\bar{u}(x)\gamma_\nu c(x)\,\bar{d}(x)\gamma_\mu c(x)  \Big] \, ,\nonumber\\
J_{{D}^*{D}_s^*,L,\mu\nu}(x)&=&\frac{1}{\sqrt{2}}\Big[\bar{q}(x)\gamma_\mu c(x)\, \bar{s}(x)\gamma_\nu c(x)-\bar{q}(x)\gamma_\nu c(x)\, \bar{s}(x)\gamma_\mu c(x) \Big] \, ,\nonumber\\
J_{{D}^*{D}_s^*,H,\mu\nu}(x)&=&\frac{1}{\sqrt{2}}\Big[\bar{q}(x)\gamma_\mu c(x)\, \bar{s}(x)\gamma_\nu c(x)+\bar{q}(x)\gamma_\nu c(x)\, \bar{s}(x)\gamma_\mu c(x) \Big] \, ,  \nonumber\\
J_{{D}^*_s{D}_s^*,L,\mu\nu}(x)&=&\frac{1}{\sqrt{2}}\Big[\bar{s}(x)\gamma_\mu c(x)\, \bar{s}(x)\gamma_\nu c(x)-\bar{s}(x)\gamma_\nu c(x)\, \bar{s}(x)\gamma_\mu c(x) \Big] \, ,
\nonumber\\
J_{{D}^*_s{D}_s^*,H,\mu\nu}(x)&=&\frac{1}{\sqrt{2}}\Big[\bar{s}(x)\gamma_\mu c(x)\, \bar{s}(x)\gamma_\nu c(x)+\bar{s}(x)\gamma_\nu c(x)\, \bar{s}(x)\gamma_\mu c(x) \Big] \, ,
\end{eqnarray}
and $q=u$, $d$. We construct  the color-singlet-color-singlet type local four-quark currents $J(x)$ and $J_\mu(x)$ to interpolate the scalar and axialvector tetraquark molecular states, respectively, and add the subscripts $L$ and $H$ to distinguish the lighter and heavier  states in the same doublet due to the mixing effects, as direct calculations indicate that there exists such a tendency. In fact, the tensor currents $J_{L,\mu\nu}(x)$ and $J_{H,\mu\nu}(x)$ couple potentially to the axialvector and tensor tetraquark molecular states, respectively, we also add the subscripts $L$ and $H$ to distinguish the lighter and heavier  states in the same doublet.
The currents $J_{D{D}^*,L,\mu}(x)$ ($J_{D_1D_0^*,H,\mu}(x)$, $J_{{D}^*{D}^*,L,\mu\nu}(x)$) and $J_{D{D}^*,H,\mu}(x)$ ($J_{D_1D_0^*,L,\mu}(x)$, $J_{{D}^*{D}^*,H,\mu\nu}(x)$) have the isospin $(I,I_3)=(0,0)$ and $(1,0)$, respectively, though they have the same constituents, we can distinguish them therefore the molecular states  according to the isospins; while the currents $J_{D{D}_s^*,L,\mu}(x)$ ($J_{D_{s1}D_{0}^*,H,\mu}(x)$, $J_{{D}^*{D}_s^*,L,\mu\nu}(x)$) and $J_{D{D}_s^*,H,\mu}(x)$ ($J_{D_{s1}D_{0}^*,L,\mu}(x)$, $J_{{D}^*{D}_s^*,H,\mu\nu}(x)$) have the isospin $I=\frac{1}{2}$. Under parity transform $\widehat{P}$, the current  operators have the  properties,
\begin{eqnarray}\label{J-parity}
\widehat{P} J(x)\widehat{P}^{-1}&=&+J(\tilde{x}) \, , \nonumber\\
\widehat{P} J_\mu(x)\widehat{P}^{-1}&=&-J^\mu(\tilde{x}) \, , \nonumber\\
\widehat{P} J_{L,\mu\nu}(x)\widehat{P}^{-1}&=&-J_{L,}{}^{\mu\nu}(\tilde{x}) \, , \nonumber\\
\widehat{P} J_{H,\mu\nu}(x)\widehat{P}^{-1}&=&+J_{H,}{}^{\mu\nu}(\tilde{x}) \, ,
\end{eqnarray}
where the coordinates $x^\mu=(t,\vec{x})$ and $\tilde{x}^\mu=(t,-\vec{x})$.
We can rewrite Eq.\eqref{J-parity} in  more explicit form,
\begin{eqnarray}
\widehat{P} J_i(x)\widehat{P}^{-1}&=&+J_i(\tilde{x}) \, ,\nonumber\\
\widehat{P} J_{L,0i}(x)\widehat{P}^{-1}&=&+J_{L,0i}(\tilde{x}) \, , \nonumber\\
\widehat{P} J_{H,ij}(x)\widehat{P}^{-1}&=&+J_{H,ij}(\tilde{x}) \, ,
\end{eqnarray}
\begin{eqnarray}
\widehat{P} J_0(x)\widehat{P}^{-1}&=&-J_0(\tilde{x}) \, , \nonumber\\
\widehat{P} J_{L,ij}(x)\widehat{P}^{-1}&=&-J_{L,ij}(\tilde{x}) \, , \nonumber\\
\widehat{P} J_{H,0i}(x)\widehat{P}^{-1}&=&-J_{H,0i}(\tilde{x}) \, ,
\end{eqnarray}
where the coordinate indexes $i$, $j=1$, $2$, $3$, and we can see clearly that there are both positive and negative components, and they couple potentially to the axialvector/tensor and pseudoscalar/vector tetraquark molecular states, respectively.  At the hadron side, we separate the contributions of the axialvector/tensor tetraquark molecular states explicitly by choosing the suitable tensor structures.

Now, let us  insert  a complete set of intermediate hadronic states with
the same quantum numbers as the currents  $J(x)$, $J_\mu(x)$ and $J_{\mu\nu}(x)$  into the
correlation functions  $\Pi(p)$, $\Pi_{\mu\nu}(p)$ and $\Pi_{\mu\nu\alpha\beta}(p)$ respectively  to obtain the hadronic representation
\cite{SVZ79-1,SVZ79-2,Reinders85}, and isolate the ground state doubly-charmed scalar/axialvector/tensor   tetraquark molecule contributions,
\begin{eqnarray}
\Pi(p)&=&\frac{\lambda_{T}^2}{M_{T}^2-p^2} +\cdots \nonumber\\
&=&\Pi_{T}(p^2)+\cdots \, ,\nonumber
\end{eqnarray}
\begin{eqnarray}
\Pi_{\mu\nu}(p)&=&\frac{\lambda_{T}^2}{M_{T}^2-p^2}\left( -g_{\mu\nu}+\frac{p_{\mu}p_{\nu}}{p^2}\right) +\cdots \nonumber\\
&=&\Pi_{T}(p^2)\left( -g_{\mu\nu}+\frac{p_{\mu}p_{\nu}}{p^2}\right)+\cdots \, ,\nonumber
\end{eqnarray}
\begin{eqnarray}
\Pi_{L,\mu\nu\alpha\beta}(p)&=&\frac{\lambda_{ T}^2}{M_{T}^2\left(M_{T}^2-p^2\right)}\left(p^2g_{\mu\alpha}g_{\nu\beta} -p^2g_{\mu\beta}g_{\nu\alpha} -g_{\mu\alpha}p_{\nu}p_{\beta}-g_{\nu\beta}p_{\mu}p_{\alpha}+g_{\mu\beta}p_{\nu}p_{\alpha}+g_{\nu\alpha}p_{\mu}p_{\beta}\right) \nonumber\\
&&+\frac{\lambda_{ T^-}^2}{M_{T^-}^2\left(M_{T^-}^2-p^2\right)}\left( -g_{\mu\alpha}p_{\nu}p_{\beta}-g_{\nu\beta}p_{\mu}p_{\alpha}+g_{\mu\beta}p_{\nu}p_{\alpha}+g_{\nu\alpha}p_{\mu}p_{\beta}\right) +\cdots  \nonumber\\
&=&\widetilde{\Pi}_{T}(p^2)\left(p^2g_{\mu\alpha}g_{\nu\beta} -p^2g_{\mu\beta}g_{\nu\alpha} -g_{\mu\alpha}p_{\nu}p_{\beta}-g_{\nu\beta}p_{\mu}p_{\alpha}+g_{\mu\beta}p_{\nu}p_{\alpha}+g_{\nu\alpha}p_{\mu}p_{\beta}\right) \nonumber\\
&&+\widetilde{\Pi}_{T}^{-}(p^2)\left( -g_{\mu\alpha}p_{\nu}p_{\beta}-g_{\nu\beta}p_{\mu}p_{\alpha}+g_{\mu\beta}p_{\nu}p_{\alpha}+g_{\nu\alpha}p_{\mu}p_{\beta}\right) \, ,\nonumber
\end{eqnarray}
\begin{eqnarray}
\Pi_{H,\mu\nu\alpha\beta}(p)&=&\frac{\lambda_{ T}^2}{M_{T}^2-p^2}\left( \frac{\widetilde{g}_{\mu\alpha}\widetilde{g}_{\nu\beta}+\widetilde{g}_{\mu\beta}\widetilde{g}_{\nu\alpha}}{2}-\frac{\widetilde{g}_{\mu\nu}\widetilde{g}_{\alpha\beta}}{3}\right) +\cdots \, \, , \nonumber \\
&=&\Pi_{T}(p^2)\left( \frac{\widetilde{g}_{\mu\alpha}\widetilde{g}_{\nu\beta}+\widetilde{g}_{\mu\beta}\widetilde{g}_{\nu\alpha}}{2}-\frac{\widetilde{g}_{\mu\nu}\widetilde{g}_{\alpha\beta}}{3}\right) +\cdots\, ,
\end{eqnarray}
where the  pole residues   $\lambda_{T}$ and $\lambda_{T^{-}}$ are defined by
\begin{eqnarray}
\langle 0|J(0)|T_{cc}(p)\rangle &=&\lambda_{T}\, ,\nonumber\\
\langle 0|J_\mu(0)|T_{cc}(p)\rangle &=&\lambda_{T}\,\varepsilon_\mu\, ,\nonumber\\
\langle 0|J_{L,\mu\nu}(0)|T_{cc}(p)\rangle &=& \frac{\lambda_{T}}{M_{T}} \, \varepsilon_{\mu\nu\alpha\beta} \, \varepsilon^{\alpha}p^{\beta}\, , \nonumber\\
\langle 0|J_{L,\mu\nu}(0)|T_{cc}^{-}(p)\rangle &=&\frac{\lambda_{T^-}}{M_{T^-}} \left(\varepsilon_{\mu}p_{\nu}-\varepsilon_{\nu}p_{\mu} \right)\, , \nonumber\\
  \langle 0|J_{H,\mu\nu}(0)|T_{cc}(p)\rangle &=& \lambda_{T}\, \varepsilon_{\mu\nu} \, ,
\end{eqnarray}
the  $\varepsilon_{\mu}/\varepsilon_{\mu\nu}$  are the polarization vectors of the doubly-charmed axialvector/tensor tetraquark molecular states.

In the present work, we accomplish  the operator product expansion  up to the vacuum condensates of dimension $10$  and take  account of the vacuum condensates $\langle\bar{q}q\rangle$, $\langle\frac{\alpha_{s}GG}{\pi}\rangle$, $\langle\bar{q}g_{s}\sigma Gq\rangle$, $\langle\bar{q}q\rangle^2$,
$\langle\bar{q}q\rangle \langle\frac{\alpha_{s}GG}{\pi}\rangle$,  $\langle\bar{q}q\rangle  \langle\bar{q}g_{s}\sigma Gq\rangle$,
$\langle\bar{q}g_{s}\sigma Gq\rangle^2$ and $\langle\bar{q}q\rangle^2 \langle\frac{\alpha_{s}GG}{\pi}\rangle$ with the assumption of vacuum saturation in a consistent way \cite{WZG-Saturation},  where  $q=u$, $d$ or $s$. It is straightforward but tedious to accomplish  the operator product expansion, for the technical details, one can consult
Refs.\cite{WangHuangTao,Wang-tetraquark-QCDSR-1,Wang-molecule-QCDSR-1,Wang-molecule-QCDSR-2}.
 Moreover, we neglect the small masses of the $u$ and $d$ quarks, and take account of  the terms proportional to  $m_s$ considering the light flavor $SU(3)$ mass-breaking effects.

Then we match  the hadron side with the QCD  side of the correlation functions $\Pi_{T}(p^2)$ and $p^2\widetilde{\Pi}_{T}(p^2)$ below the continuum thresholds  $s_0$ and accomplish  Borel transform  in regard  to
the variable  $P^2=-p^2$ to obtain  the  QCD sum rules:
\begin{eqnarray}\label{QCDSR}
\lambda^2_{T}\, \exp\left(-\frac{M^2_{T}}{T^2}\right)= \int_{4m_c^2}^{s_0} ds\, \rho_{QCD}(s) \, \exp\left(-\frac{s}{T^2}\right) \, ,
\end{eqnarray}
 the  QCD spectral densities $\rho_{QCD}(s)$ are available via contacting us via E-mail.
We  differentiate  Eq.\eqref{QCDSR} in regard  to  $\tau=\frac{1}{T^2}$,  and obtain the QCD sum rules for the masses of the    scalar/axialvector/tensor   doubly-charmed tetraquark molecular states $T_{cc}$ without strange, with strange and with doubly-strange,
 \begin{eqnarray}\label{QCDSR-diff}
 M^2_{T}&=& -\frac{\int_{4m_c^2}^{s_0} ds\frac{d}{d \tau}\rho_{QCD}(s)\exp\left(-\tau s \right)}{\int_{4m_c^2}^{s_0} ds \rho_{QCD}(s)\exp\left(-\tau s\right)}\mid_{\tau=\frac{1}{T^2}}\, .
\end{eqnarray}
We acquire the masses and pole residues by solving the coupled equations \eqref{QCDSR}-\eqref{QCDSR-diff}.

\section{Numerical results and discussions}
 We take  the conventional  values of the  vacuum condensates
$\langle\bar{q}q \rangle=-(0.24\pm 0.01\, \rm{GeV})^3$,  $\langle\bar{s}s \rangle=(0.8\pm0.1)\langle\bar{q}q \rangle$,
 $\langle\bar{q}g_s\sigma G q \rangle=m_0^2\langle \bar{q}q \rangle$, $\langle\bar{s}g_s\sigma G s \rangle=m_0^2\langle \bar{s}s \rangle$,
$m_0^2=(0.8 \pm 0.1)\,\rm{GeV}^2$, $\langle \frac{\alpha_s
GG}{\pi}\rangle=0.012\pm0.004\,\rm{GeV}^4$     at the energy scale  $\mu=1\, \rm{GeV}$
\cite{SVZ79-1,SVZ79-2,Reinders85,Colangelo-Review}, and  take the $\overline{MS}$ masses $m_{c}(m_c)=(1.275\pm0.025)\,\rm{GeV}$
 and $m_s(\mu=2\,\rm{GeV})=(0.095\pm0.005)\,\rm{GeV}$
 from the Particle Data Group \cite{PDG}.
In addition,  we take account of
the energy-scale dependence of  the quark condensates, mixed quark condensates and $\overline{MS}$ masses \cite{Narison-mix},
 \begin{eqnarray}
 \langle\bar{q}q \rangle(\mu)&=&\langle\bar{q}q\rangle({\rm 1 GeV})\left[\frac{\alpha_{s}({\rm 1 GeV})}{\alpha_{s}(\mu)}\right]^{\frac{12}{33-2n_f}}\, , \nonumber\\
 \langle\bar{s}s \rangle(\mu)&=&\langle\bar{s}s \rangle({\rm 1 GeV})\left[\frac{\alpha_{s}({\rm 1 GeV})}{\alpha_{s}(\mu)}\right]^{\frac{12}{33-2n_f}}\, , \nonumber\\
 \langle\bar{q}g_s \sigma Gq \rangle(\mu)&=&\langle\bar{q}g_s \sigma Gq \rangle({\rm 1 GeV})\left[\frac{\alpha_{s}({\rm 1 GeV})}{\alpha_{s}(\mu)}\right]^{\frac{2}{33-2n_f}}\, ,\nonumber\\
  \langle\bar{s}g_s \sigma Gs \rangle(\mu)&=&\langle\bar{s}g_s \sigma Gs \rangle({\rm 1 GeV})\left[\frac{\alpha_{s}({\rm 1 GeV})}{\alpha_{s}(\mu)}\right]^{\frac{2}{33-2n_f}}\, ,\nonumber\\
m_c(\mu)&=&m_c(m_c)\left[\frac{\alpha_{s}(\mu)}{\alpha_{s}(m_c)}\right]^{\frac{12}{33-2n_f}} \, ,\nonumber\\
m_s(\mu)&=&m_s({\rm 2GeV} )\left[\frac{\alpha_{s}(\mu)}{\alpha_{s}({\rm 2GeV})}\right]^{\frac{12}{33-2n_f}}\, ,\nonumber\\
\alpha_s(\mu)&=&\frac{1}{b_0t}\left[1-\frac{b_1}{b_0^2}\frac{\log t}{t} +\frac{b_1^2(\log^2{t}-\log{t}-1)+b_0b_2}{b_0^4t^2}\right]\, ,
\end{eqnarray}
  where $t=\log \frac{\mu^2}{\Lambda^2}$, $b_0=\frac{33-2n_f}{12\pi}$, $b_1=\frac{153-19n_f}{24\pi^2}$, $b_2=\frac{2857-\frac{5033}{9}n_f+\frac{325}{27}n_f^2}{128\pi^3}$,  $\Lambda=213\,\rm{MeV}$, $296\,\rm{MeV}$  and  $339\,\rm{MeV}$ for the quark flavors  $n_f=5$, $4$ and $3$, respectively  \cite{PDG}.
There are $u$, $d$, $s$ and $c$ quarks, we choose the quark flavor numbers $n_f=4$, and evolve the QCD spectral densities $\rho_{QCD}(s)$ to the suitable energy scales $\mu$ to extract the molecule masses.

In the hidden-heavy $Q\bar{Q}q\bar{q}^\prime$ systems and doubly-heavy $QQ\bar{q}\bar{q}^\prime$ systems,  we can introduce the effective heavy quark masses $\mathbb{M}_Q$ and divide the tetraquark (molecular) states into both the heavy degrees of freedoms and light degrees of freedoms. We neglect the small $u$ and $d$ quark masses, then obtain the heavy degrees of freedoms  $2{\mathbb{M}}_Q$ and light degrees of freedoms $\mu=\sqrt{M^2_{X/Y/Z/T}-(2{\mathbb{M}}_Q)^2}$ \cite{QQ-QCDSR-WZG-APPB,QQ-QCDSR-WZG-EPJC,Wang-tetraquark-QCDSR-1,Wang-molecule-QCDSR-1,Wang-molecule-QCDSR-2,Wang-tetraquark-QCDSR-2,Wang-tetraquark-QCDSR-3,Wang-tetraquark-QCDSR-4}.  We can also  introduce the effective $s$-quark mass $\mathbb{M}_s$ according to the light flavor $SU(3)$ breaking effects, then the light degrees of freedoms $\mu=\sqrt{M^2_{X/Y/Z/T}-(2{\mathbb{M}}_c)^2}-k\,\mathbb{M}_s$ with $k=0$, $1$, $2$ \cite{Wang-molecule-QCDSR-3,WZG-XQ-penta-mole}. In the present work, we choose the effective $c/s$-quark masses $\mathbb{M}_c=1.82\,\rm{GeV}$ and $\mathbb{M}_s=0.2\,\rm{GeV}$, and use the modified energy scale formula  $\mu=\sqrt{M^2_{X/Y/Z/T}-(2{\mathbb{M}}_c)^2}-k\,\mathbb{M}_s$ to enhance the pole contributions and improve the convergent behavior of the operator product expansion.

  The energy gaps between the ground states and first radial excited states are about $0.5\sim 0.6\,\rm{GeV}$ for the traditional heavy mesons and quarkonia \cite{PDG}, we tentatively take the constraint $\sqrt{s_0}=M_{T}+0.4\sim 0.6\,\rm{GeV}$, as there are two color-singlet clusters, each cluster has the same quantum numbers as that of the corresponding heavy meson,  and search for  the continuum threshold parameters $s_0$ and Borel parameters $T^2$ to satisfy
pole dominance at the hadron  side and  convergence of the operator product expansion at the QCD side  via trial  and error.

Eventually, we obtain the Borel parameters, continuum threshold parameters, energy scales of the QCD spectral densities,  pole contributions, and the contributions of the highest dimensional  vacuum condensates, which are presented  clearly  in Table \ref{BorelP}. In the Table, we use the notations $D$, $D^*$, $\cdots$ to represent the color-singlet constituents in the molecular states with the same quantum numbers as the mesons $D$, $D^*$, $\cdots$.
From the Table,  we can see clearly  that the pole contributions or ground state contributions, which are defined by
\begin{eqnarray}
{\rm{pole}}&=&\frac{\int_{4m_{c}^{2}}^{s_{0}}ds\rho_{QCD}\left(s\right)\exp\left(-\frac{s}{T^{2}}\right)} {\int_{4m_{c}^{2}}^{\infty}ds\rho_{QCD}\left(s\right)\exp\left(-\frac{s}{T^{2}}\right)}\, ,
\end{eqnarray}
 are  larger than $(40-60)\%$ at the hadron side, the central values are larger than $50\%$, the elementary  criterion of the pole dominance is  satisfied very good.
 Moreover, the contributions of the vacuum condensates of dimension $10$ are $|D(10)|< 1 \%$ or $\ll 1\%$ at the QCD side, where the contributions from the QCD spectral densities $\rho_{QCD,n}(s)$ having the vacuum condensates of dimension $n$ are defined by
 \begin{eqnarray}
D(n)&=&\frac{\int_{4m_{c}^{2}}^{s_{0}}ds\rho_{QCD,n}(s)\exp\left(-\frac{s}{T^{2}}\right)}
{\int_{4m_{c}^{2}}^{s_{0}}ds\rho_{QCD}\left(s\right)\exp\left(-\frac{s}{T^{2}}\right)}\, ,
\end{eqnarray}
the convergent behaviors  of the operator product  expansion are  very good, the other elementary  criterion  is also satisfied very good.

Now we take  account of  all the uncertainties of the input parameters,  and obtain the masses and pole residues of the  scalar/axialvector/tensor   doubly-charmed  tetraquark molecular  states without strange, with strange and with doubly-strange, which are presented  explicitly in Table \ref{mass-Table}. From the Tables \ref{BorelP}--\ref{mass-Table}, we can see clearly that the modified energy scale formula $\mu=\sqrt{M^2_{X/Y/Z/T}-(2{\mathbb{M}}_c)^2}-k\,\mathbb{M}_s$ is well satisfied.

\begin{figure}
\centering
\includegraphics[totalheight=6cm,width=7cm]{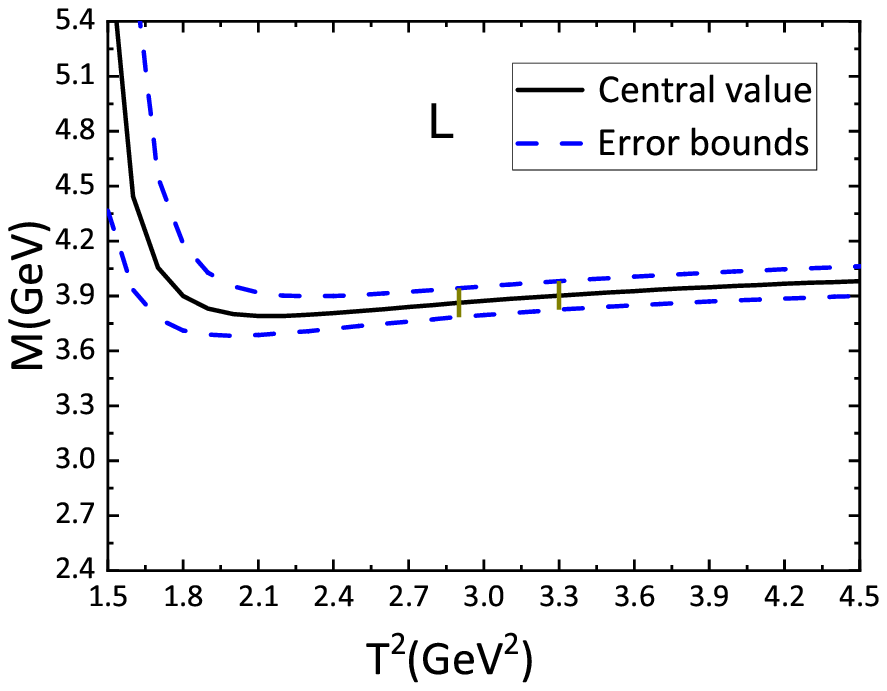}
\includegraphics[totalheight=6cm,width=7cm]{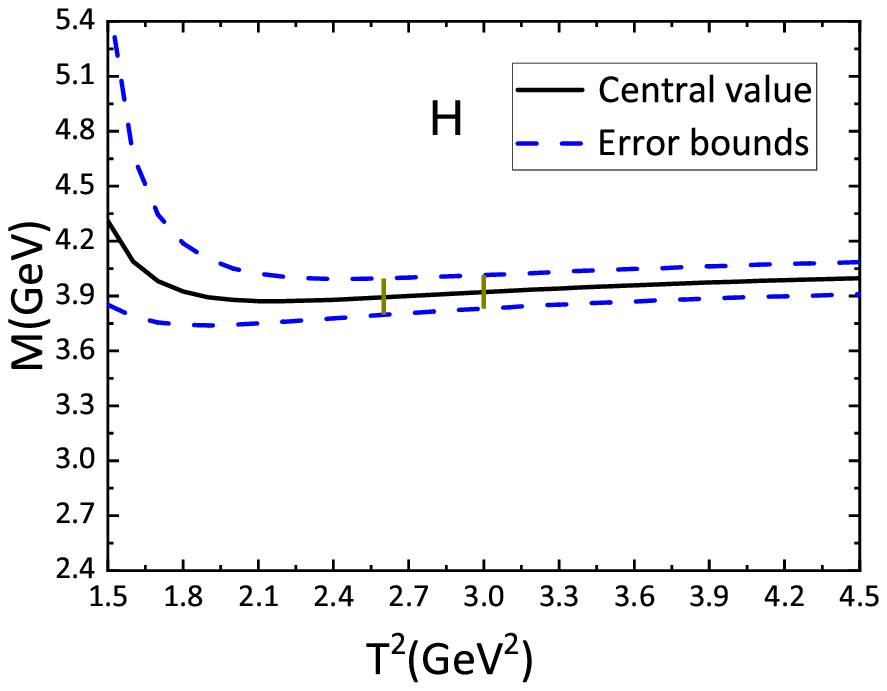}
  \caption{ The masses  with variations of the  Borel parameters for the axialvector tetraquark  molecular states, where the $L$ and
$H$ denote the lighter and heavier $DD^{*}$ states,  respectively.   } \label{mass-Borel}
\end{figure}

 In  Fig.\ref{mass-Borel}, as an example, we plot the masses of the  axialvector doubly-charmed tetraquark molecular states $(D^{*}D -DD^{*})_L$ and
$(D^{*}D +DD^{*})_H$  with variations of the Borel parameters at much larger ranges than the Borel widows. From the figure, we can see clearly that there appear very flat platforms in the Borel windows,  the regions between the two short perpendicular lines.    The uncertainties  originate from the Borel parameters are rather small, and we expect to make reliable predictions. In the present work, we choose the uniform constraints $\sqrt{s_0}= M_T+0.50\sim0.55\pm 0.10\,\rm{GeV}$ and $\mu=\sqrt{M^2_{X/Y/Z/T}-(2{\mathbb{M}}_c)^2}-k\,\mathbb{M}_s$, uniform pole contributions $(40-60)\%$, uniform convergent behaviors $|D(10)|< 1 \%$ or $\ll 1\%$, and very flat Borel platforms, and we expect to make robust predictions.

There exist both a lighter state and a heavier state for the $cc\bar{u}\bar{d}$ and $cc\bar{q}\bar{s}$    tetraquark molecular states, the lighter state
$(D^{*}D -DD^{*})_L$  with the isospin $(I,I_3)=(0,0)$  has a  mass $3.88\pm0.11\,\rm{GeV}$, which is in excellent agreement with the mass of the doubly-charmed tetraquark candidate $T_{cc}^+$ from the LHCb collaboration \cite{Tcc:talk,LHCb-Tcc,LHCb-Tcc-detail}, and supports assigning the $T_{cc}^+$ to be the $(D^{*}D -DD^{*})_L$ molecular state, as the $T_{cc}^+$ has the isospin $I=0$. In other words, the exotic state $T_{cc}^+$ maybe have  a $(D^{*}D -DD^{*})_L$ Fock component.  The heavier state $(D^{*}D +DD^{*})_H$ with the isospin $(I,I_3)=(1,0)$ has a  mass  $3.90\pm0.11\,\rm{GeV}$, the central value lies slightly above the $DD^*$ threshold, the strong decays to the final states $DD\pi$ are  kinematically allowed but with small phase-space. While in the scenario of the tetraquark states, the doubly-charmed tetraquark states with the isospin $(I,I_3)=(1,0)$ and $(0,0)$ have degenerated masses $3.90\pm0.09\,\rm{GeV}$ \cite{QQ-QCDSR-WZG-APPB,QQ-QCDSR-WZG-EPJC}.

If we choose the same input parameters, the $DD^*$ molecular state with the isospin $I=1$ has slightly larger mass than the corresponding molecule with the isospin $I=0$, it is  indeed that the isoscalar $DD^*$ molecular state is lighter.  To reduce  systematic uncertainties and make more reliable predictions, we choose the same pole contributions to acquire all the molecule masses, irrespective of the isospins $I=1$, $\frac{1}{2}$ and $0$, and tentatively assign the LHCb's $T_{cc}^+$ as the molecular state with the isospin $I=0$, because  exploring the $DD$ mass spectrum  disfavors interpreting  the  $T_{cc}^+$ as the isovector state \cite{LHCb-Tcc,LHCb-Tcc-detail}.

From Table \ref{mass-Table}, we can see clearly that there exists a small mass gap between the centroids  of the isoscalar $(D^{*}D -DD^{*})_L$ and isovector
$(D^{*}D +DD^{*})_H$ states, about $0.02\,\rm{GeV}$, while the mass gap between the centroids  of the two states $(D_s^*D -D_sD^*)_L$ and $(D_s^*D +D_sD^*)_H$ with the isospin $I=\frac{1}{2}$ is about $0.01\,\rm{GeV}$. The small but finite mass gaps maybe originate from the spin-spin and tensor interactions between the constituent quarks, as the isospin breaking effect of the $u$ and $d$ quarks is about $\delta m=3.4\,\rm{MeV}$ at the energy scale $\mu=1\,\rm{GeV}$ \cite{PDG}, which cannot account for the mass gaps $0.02\,\rm{GeV}$ and $0.01\,\rm{GeV}$. For the molecular states $(D_0^*D_1 +D_1D_0^*)_L$, $(D_0^*D_1 -D_1D_0^*)_H$,
$(D_{0}^*D_{s1} +D_{s0}^*D_1)_L$ and
$(D_{0}^*D_{s1} -D_{s0}^*D_1)_H$, there exists  a  P-wave in the color-singlet constituents, the P-wave is embodied implicitly in the underlined $\underline{\gamma_5}$ in the scalar currents $\bar{q}i\gamma_5 \underline{\gamma_5}c$, $\bar{s}i\gamma_5 \underline{\gamma_5}c$ and axialvector currents  $\bar{q}\gamma_\mu \underline{\gamma_5} c$, $\bar{s}\gamma_\mu \underline{\gamma_5} c$, as multiplying $\gamma_5$ to the pseudoscalar  currents $\bar{q}i\gamma_5 c$, $\bar{s}i\gamma_5 c$ and vector currents  $\bar{q}\gamma_\mu  c$, $\bar{s}\gamma_\mu  c$ changes their parity. We should introduce the spin-orbit interactions to account for the large mass gaps between the lighter and heavier states $(L,H)$, i.e. $((D_0^*D_1 +D_1D_0^*)_L, \,(D_0^*D_1 -D_1D_0^*)_H)$,
$((D_{0}^*D_{s1} +D_{s0}^*D_1)_L, \,(D_{0}^*D_{s1} -D_{s0}^*D_1)_H)$.

From Tables \ref{mass-Borel}-\ref{mass-Table}, we can see that the continuum threshold parameters  $\sqrt{s_0}= M_T+0.50\sim0.55\pm 0.10\,\rm{GeV}$, one maybe worry about the contaminations from the  two-meson or three-meson  scattering states, which contribute self-energies. In fact, the renormalized self-energies  contribute  a finite imaginary part to modify the dispersion relation, we can take  account of the finite width effects by the  simple replacement of the hadronic spectral densities,
\begin{eqnarray}
\lambda^2_{T}\delta \left(s-M^2_{T} \right) &\to& \lambda^2_{T}\frac{1}{\pi}\frac{M_{T}\Gamma_{T}(s)}{(s-M_{T}^2)^2+M_{T}^2\Gamma_{T}^2(s)}\, ,
\end{eqnarray}
where the $\Gamma_{T}(s)$ are the energy dependent widths.  All in all, we can take account of the two-meson or three-meson  scattering states reasonably by adding a finite width to the
 tetraquark (molecular) states. Direct calculations indicate that the finite widths cannot  affect  the masses $M_{T}$ significantly, and can be safely absorbed into the pole residues saving the physical widths are not large enough. For detailed discussions about this subject,  one can consult Ref.\cite{WZG-Landau}.
Furthermore, we choose the local four-quark  currents, which couple potentially to the molecular states rather than to the scattering states, although the couplings to the scattering states are unavoidable considering the same quantum numbers.  The traditional mesons are spatial extended objects and have average  spatial sizes $\sqrt{\langle r^2\rangle} \neq 0$.
In the local limit $r \to 0$,   the  $D$
and  $D^*$ mesons lose themselves and merge into color-singlet-color-singlet type tetraquark states. We expect that the $D$, $D^*$ and $T_{cc}$ mesons have average  spatial sizes of the same order, the couplings to the continuum states are small, as the overlappings  of the wave-functions are rather  small.
In the QCD sum rules, we use the  nomenclature "molecular states" in the sense of color-singlet-color-singlet type structures, and the color-singlet clusters have the same quantum numbers as the traditional mesons,   in fact, they are also compact tetraquark states according to the local currents, just like the color-antitriplet-color-triplet (diquark-antidiquark) type tetraquark states. If they are the usually called "molecular states" indeed, they should have the spatial extension about (or larger than) $1\,\rm{fm}$, which is much larger than  the size of the traditional mesons,  and it is not robust to interpolate them with  the local four-quark currents.

The decay properties of the doubly-charmed tetraquark molecular states are rather simple as the decays  can take place easily  through the Okubo-Zweig-Iizuka super-allowed fall-apart mechanism if kinematically allowed,
we can search for those scalar/axialvector/tensor doubly-charmed tetraquark molecular states without strange, with strange and with doubly-strange presented  in Table \ref{mass-Table} in
the $DD\pi $, $DD\gamma $, $DD\pi \pi$,  $DD_s\gamma$, $DD_s\pi\pi$,  $DD_s\pi\pi\pi$, $D_sD_s\gamma$,  $D_sD_s\pi\pi\pi$ invariant mass distributions  at the BESIII, LHCb, Belle II,  CEPC, FCC, ILC in the future. For example, the production cross sections  of the tetraquark state $cc\bar{u}\bar{d}$ with the spin-parity $J^P=1^+$ were  explored before the observation of the $T_{cc}^+$ by the LHCb collaboration \cite{Product-QiQi}.

\begin{table}
\begin{center}
\begin{tabular}{|c|c|c|c|c|c|c|c|c|}\hline\hline
$T_{cc}$                           &Isospin        &$T^2(\rm{GeV}^2)$   & $\sqrt{s_0}(\rm GeV) $   &$\mu(\rm{GeV})$  &pole          &$|D(10)|$        \\ \hline

$D^{*}D^{*}$                       &1              & $2.8-3.2$          & $4.55\pm0.10$            &$1.7$            &$(41-61)\%$   &$<1\%$            \\

$D_{s}^*D^{*}$                     &$\frac{1}{2}$  & $2.9-3.3$          & $4.65\pm0.10$            &$1.7$            &$(42-62)\%$   &$\ll1\%$        \\

$D_{s}^*D_{s}^*$                   & 0             & $3.2-3.5$          & $4.80\pm0.10$            &$1.8$           &$(42-61)\%$    &$\ll1\%$       \\

$D^{*}D -DD^{*}$                   &0              & $2.9-3.3$          & $4.45\pm0.10$            &$1.4$            &$(42-62)\%$   &$\ll1\%$     \\

$D^{*}D +DD^{*}$                   &1              & $2.6-3.0$          & $4.40\pm0.10$            &$1.4$            &$(42-63)\%$   &$\ll1\%$   \\

$D_s^*D -D_sD^*$                   &$\frac{1}{2}$  & $3.0-3.4$          & $4.50\pm0.10$            &$1.5$            &$(40-62)\%$   &$<1\%$   \\

$D_s^*D +D_sD^*$                   &$\frac{1}{2}$  & $2.9-3.3$          & $4.50\pm0.10$            &$1.5$            &$(40-60)\%$   &$\ll1\%$ \\

$D_s^*D_s$                         & 0             & $3.0-3.4$          & $4.60\pm0.10$            &$1.5$            &$(41-63)\%$   &$\ll1\%$  \\

$D_0^*D_1 -D_1D_0^*$               & 0             & $5.6-7.0$          & $6.35\pm0.10$            &$4.6$            &$(41-60)\%$   &$\ll1\%$   \\

$D_0^*D_1 +D_1D_0^*$               & 1             & $4.7-6.1$          & $5.90\pm0.10$            &$4.0$            &$(42-61)\%$   &$\ll1\%$   \\

$D_{0}^*D_{s1} -D_{s0}^*D_1$       &$\frac{1}{2}$  & $5.8-7.2$          & $6.50\pm0.10$            &$4.6$             &$(43-60)\%$  &$<1\%$   \\

$D_{0}^*D_{s1} +D_{s0}^*D_1$       &$\frac{1}{2}$  & $4.7-6.1$          & $6.05\pm0.10$            &$4.0$             &$(42-62)\%$  &$<1\%$   \\

$D_{s1}D_{s0}^*$                   & 0             & $4.9-6.3$          & $6.20\pm0.10$            &$4.0$             &$(43-61)\%$  &$<1\%$   \\

$D^{*}D^{*}-D^{*}D^{*}$            &0              & $3.2-3.6$          & $4.55\pm0.10$            &$1.7$             &$(42-61)\%$  &$<1\%$ \\

$D^{*}D^{*}+D^{*}D^{*}$            &1              & $3.0-3.4$          & $4.55\pm0.10$            &$1.7$             &$(41-60)\%$  &$<1\%$  \\

$D_{s}^*D^{*}-D_{s}^{*}D^*$        &$\frac{1}{2}$  & $3.3-3.7$          & $4.65\pm0.10$            &$1.7$             &$(40-59)\%$  &$\ll1\%$ \\

$D_{s}^*D^{*}+D_{s}^{*}D^*$        &$\frac{1}{2}$  & $3.1-3.5$          & $4.65\pm0.10$            &$1.7$             &$(42-61)\%$  &$<1\%$   \\

$D_{s}^*D_{s}^*-D_{s}^*D_{s}^*$    & 0             & $3.6-4.0$          & $4.80\pm0.10$            &$1.8$             &$(40-60)\%$  &$\ll1\%$ \\

$D_{s}^*D_{s}^*+D_{s}^*D_{s}^*$    & 0             & $3.4-3.9$          & $4.80\pm0.10$            &$1.8$             &$(41-61)\%$  &$<1\%$    \\
\hline\hline
\end{tabular}
\end{center}
\caption{ The Borel parameters, continuum threshold parameters, energy scales of the QCD spectral densities,  pole contributions, and the contributions of the vacuum condensates of dimension $10$  for the ground state doubly-charmed  axialvector tetraquark molecular states. }\label{BorelP}
\end{table}

\begin{table}
\begin{center}
\begin{tabular}{|c|c|c|c|c|c|c|c|c|}\hline\hline
$T_{cc}$                           &Isospin          & $M_T (\rm{GeV})$   & $\lambda_T (\rm{GeV}^5) $        \\ \hline

$D^{*}D^{*}$                       & 1               & $4.04\pm0.11$      &$(4.99\pm0.66)\times 10^{-2}$     \\

$D_{s}^*D^{*}$                     &$\frac{1}{2}$    & $4.12\pm0.10$      & $(5.74\pm0.78)\times 10^{-2}$      \\

$D_{s}^*D_{s}^*$                   & 0               & $4.22\pm0.10$      & $(7.46\pm0.89)\times 10^{-2}$    \\

$D^{*}D -DD^{*}$                   & 0               & $3.88\pm0.11$      & $(1.92\pm0.29)\times 10^{-2}$    \\

$D^{*}D +DD^{*}$                   & 1               & $3.90\pm0.11$      & $(1.50\pm0.22)\times 10^{-2}$     \\

$D_s^*D -D_sD^*$                   &$\frac{1}{2}$    & $3.97\pm0.10$      & $(2.40\pm0.41)\times 10^{-2}$     \\

$D_s^*D +D_sD^*$                   &$\frac{1}{2}$    & $3.98\pm0.11$      & $(2.06\pm0.30)\times 10^{-2}$      \\

$D_s^*D_s$                         & 0               & $4.10\pm0.12$      & $(2.31\pm0.45)\times10^{-2}$      \\

$D_0^*D_1 -D_1D_0^*$               & 0               & $5.79\pm0.15$      & $(2.13\pm0.19)\times 10^{-1}$     \\

$D_0^*D_1 +D_1D_0^*$               & 1               & $5.37\pm0.13$      & $(1.30\pm0.11)\times 10^{-1}$    \\

$D_{0}^*D_{s1} -D_{s0}^*D_1$       &$\frac{1}{2}$    & $5.93\pm0.27$      & $(2.80\pm0.33)\times 10^{-1}$    \\

$D_{0}^*D_{s1} +D_{s0}^*D_1$       &$\frac{1}{2}$    & $5.54\pm0.20$      & $(1.51\pm0.16)\times 10^{-1}$     \\

$D_{s1}D_{s0}^*$                   & 0               & $5.67\pm0.27$      & $(1.77\pm0.27)\times 10^{-1}$     \\

$D^{*}D^{*}-D^{*}D^{*}$            & 0               & $4.00\pm0.11$      & $(2.47\pm0.32)\times 10^{-2}$    \\

$D^{*}D^{*}+D^{*}D^{*}$            & 1               & $4.02\pm0.11$      & $(2.83\pm0.30)\times 10^{-2}$     \\

$D_{s}^*D^{*}-D_{s}^{*}D^*$        &$\frac{1}{2}$    & $4.08\pm0.10$      & $(2.81\pm0.40)\times 10^{-2}$     \\

$D_{s}^*D^{*}+D_{s}^{*}D^*$        &$\frac{1}{2}$    & $4.10\pm0.11$      & $(3.19\pm0.44)\times 10^{-2}$      \\

$D_{s}^*D_{s}^*-D_{s}^*D_{s}^*$    & 0               & $4.19\pm0.09$      & $(3.49\pm0.49)\times 10^{-2}$      \\

$D_{s}^*D_{s}^*+D_{s}^*D_{s}^*$    & 0               & $4.20\pm0.10$      & $(4.00\pm0.53)\times 10^{-2}$    \\
\hline\hline
\end{tabular}
\end{center}
\caption{ The masses and pole residues of the ground state doubly-charmed  tetraquark molecular states. }\label{mass-Table}
\end{table}

\begin{figure}
\centering
\includegraphics[totalheight=7cm,width=10cm]{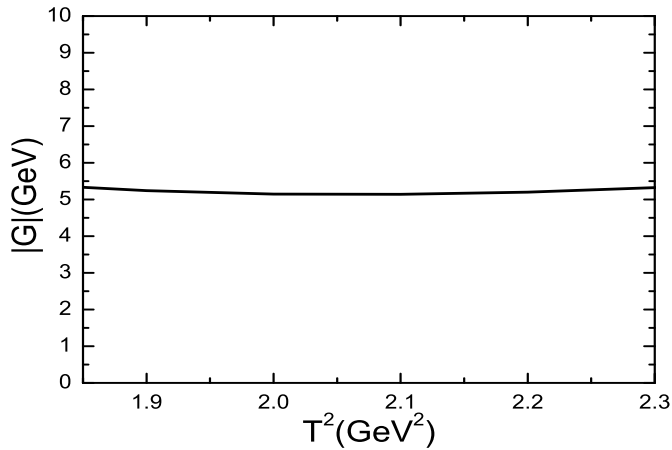}
  \caption{ The hadronic coupling constant $G$  with variations of the  Borel parameter $T^2$.   } \label{G-Borel}
\end{figure}

Now we explore the three-body strong decays $T_{cc}^+\to D^0D^{*+} \to D^0D^0 \pi^+$, $D^0D^+\pi^0$, and $T_{cc}^+\to D^+D^{*0} \to D^+D^0 \pi^0$ to acquire additional support of assigning the $T_{cc}^+$ to be the $D^{*}D -DD^{*}$ molecular state.  We write down the three-point correlation function $\Pi_{\alpha\mu}(p,q)$,
\begin{eqnarray}
\Pi_{\alpha\mu}(p,q)&=&i^2\int d^4xd^4y \, e^{ip \cdot x}e^{iq \cdot y}\, \langle 0|T\left\{J_5^{D}(x)J_\alpha^{D^*}(y)J_{\mu}^{DD^*,L}(0)\right\}|0\rangle \, ,
\end{eqnarray}
where the currents,
\begin{eqnarray}
J_5^{D}(x)&=&\bar{c}(x)i\gamma_5 u(x) \, ,\nonumber \\
J_\alpha^{D^*}(y)&=&\bar{c}(y)\gamma_\alpha d(y) \, ,
\end{eqnarray}
 interpolate the $D$ and $D^*$  mesons, respectively. At the hadron side, we isolate the ground state
contributions,
\begin{eqnarray}
\Pi_{\alpha\mu}(p,q)&=&\left\{\frac{f_{D}M_{D}^2f_{D^*}M_{D^*}\lambda_{T}G}{\sqrt{2}m_c} \frac{1}{(M_{T}^2-p^{\prime2})(M_{D}^2-p^2)(M_{D^*}^2-q^2)} \right. \nonumber\\
&&\left.+  \frac{ C_{DD^* }}{(M_{D}^2-p^{2})(M_{D^*}^2-q^2)}+\cdots \right\} \,g_{\alpha\mu}+\cdots \, .
\end{eqnarray}
\begin{eqnarray}
\langle0|J_{5}^{D}(0)|D(p)\rangle&=&\frac{f_{D}M_{D}^2}{m_c} \,\, , \nonumber \\
\langle0|J_{\alpha}^{D^*}(0)|D^*(q)\rangle&=&f_{D^*}M_{D^*}\,\xi_\alpha \,\, , \nonumber \\
\langle D(p) D^*(q)|T_{cc}(p^{\prime})\rangle&=&i \xi^*(q)\cdot \varepsilon(p^{\prime})\, \frac{G}{\sqrt{2}} \, ,
\end{eqnarray}
the $\xi_\alpha$ is polarization vector of the $D^*$ meson, the $G$ is the hadronic coupling constant in the  lagrangian,
\begin{eqnarray}
{\cal L} &=& \frac{G}{\sqrt{2}} \, T_{cc,\alpha}^{+\dagger}\left(D^0D^{*+,\alpha}- D^+D^{*0,\alpha}\right)+h.c. \, .
\end{eqnarray}
For the definition of the unknown constant $C_{DD^*}$, one can consult Refs.\cite{WZG-ZJX-EPJC,WZG-Y4660-decay}. We choose  the tensor structure $g_{\alpha\mu}$  to study the hadronic  coupling constant $G$ and  accomplish  the operator product expansion up to the vacuum condensates of dimension $10$
at the QCD side. Then we take the rigorous  quark-hadron duality below the continuum thresholds,  set $p^{\prime2}=4p^2$,  and perform the double Borel transforms with respect to the variables $P^2=-p^2$ and $Q^2=-q^2$, respectively  to obtain the QCD sum rules \cite{WZG-ZJX-EPJC,WZG-Y4660-decay},
\begin{eqnarray}
&&\frac{f_{D}M_{D}^2f_{D^*}M_{D^*}\lambda_{T}G}{4\sqrt{2}m_c } \frac{1}{\widetilde{M}_{T}^2-M_{D}^2}\left[ \exp\left(-\frac{M_{D}^2}{T_1^2} \right)-\exp\left(-\frac{\widetilde{M}_{T}^2}{T_1^2} \right)\right]\exp\left(-\frac{M_{D^*}^2}{T_2^2} \right) \nonumber\\
&&+ C_{D D^*} \exp\left(-\frac{M_{D}^2}{T_1^2} -\frac{M_{D^*}^2}{T_2^2} \right)=\int_{m_c^2}^{s^0_{D}} ds \int_{m_c^2}^{u^0_{D^*}} du  \rho_{QCD}(s,u)   \exp\left(-\frac{s}{T_1^2} -\frac{u}{T_2^2} \right)\, ,
\end{eqnarray}	
where $\widetilde{M}_{T}^2=\frac{M_{T}^2}{4}$, the  $s^0_{D}$ and $u^0_{D^*}$ are the continuum threshold parameters, the $T_1^2$ and $T^2_2$ are the Borel parameters, and the $\rho_{QCD}(s,u)$ is the QCD spectral density, which is neglected for simplicity.
Again, we set $T_1^2=T_2^2=T^2$ for simplicity \cite{WZG-Y4660-decay}, and choose the  hadronic parameters
$M_{D}=1.87\,\rm{GeV}$, $f_{D}=208\,\rm{MeV}$, $s^0_{D}=(2.4\,\rm{GeV})^2$, $M_{D^*}=2.01\,\rm{GeV}$, $f_{D^*}=263\,\rm{MeV}$, $u^0_{D^*}=(2.5\,\rm{GeV})^2$ from the QCD sum rules   \cite{WZG-decay-constant}.
The unknown parameter is fitted to be $\sqrt{2}C_{D D^*}=-0.0215\,\rm{GeV}^8$ to obtain  platform in the Borel window $T^2=(1.9-2.3)\,\rm{GeV}^2$, see Fig.\ref{G-Borel}. Finally, we  acquire  the central value,
\begin{eqnarray}
|G|&=&5.2\,\rm{GeV}\, ,
\end{eqnarray}
and $\frac{|G|}{\sqrt{2}}=3.7\,\rm{GeV}$, which is in very good agreement with the values $g_{T_{cc}D^{*+}D^0}=3658.30\,\rm{MeV}$ and $g_{T_{cc}D^{*0}D^+}=-3921.04 \,\rm{MeV}$
from the unitary method with the coupled channel effects \cite{QQ-CCA-mole-Liang}, which lead to remarkable agreement with
the experimental data on the decays $T_{cc}^+ \to D^0D^0\pi^+$ \cite{LHCb-Tcc,LHCb-Tcc-detail}. Now we can obtain the conclusion tentatively that the present predictions support assigning the $T^+_{cc}$ as the $D^{*}D -DD^{*}$ molecular state, the central value $|G|=5.2\,\rm{GeV}$ serves as a crude estimation, more detailed studies (including error analysis) of the decays of  all the molecular states shown in Table \ref{mass-Table} will be our next work.

Generally speaking,  a physical  tetraquark (molecular) state maybe have several Fock components, we can choose any current with the same quark structure as one of the Fock components to interpolate this tetraquark (molecular) state due to the non-vanishing current-hadron coupling constant.
In both the scenarios tetraquark states and molecular states, we can reproduce the mass of the $T_{cc}^+$, and therefore  obtain the conclusion tentatively that
there maybe exist a color-singlet-color-singlet type molecular state and a color-antitriplet-color-triplet type tetraquark state, which have almost degenerated  masses,  or exist one axialvector  tetraquark state with both the color-singlet-color-singlet type and color-antitriplet-color-triplet type Fock components.  The present work does not exclude assigning the $T_{cc}^+$ as the diquark-antidiquark type tetraquark state. More experimental and theoretical works are still needed before reaching  definite conclusion.

\section{Conclusion}
In the present work,  we  investigate the  scalar/axialvector/tensor   doubly-charmed tetraquark molecular states  without strange, with strange and with doubly-strange via the QCD sum rules  by  performing  the operator product expansion up to the vacuum condensates of dimension $10$ and  taking  account of all the light flavor $SU(3)$ breaking effects in a consistent way. We use the modified energy scale formula to acquire the suitable   energy scales of the QCD spectral densities, and  obtain the  masses of  the ground state  scalar/axialvector/tensor   doubly-charmed tetraquark molecular states.  The present calculations favor assigning the doubly-charmed tetraquark candidate $T^+_{cc}$ to be the lighter $D^{*}D -DD^{*}$ tetraquark molecular state with the spin-parity $J^P=1^+$ and isospin $I=0$, while the
    heavier $D^{*}D +DD^{*}$  molecular state with the spin-parity $J^P=1^+$ and $I=1$  still escapes  experimental detections, the observation of the heavier $D^{*}D +DD^{*}$  molecular state would shed light on the nature of the $T_{cc}^+$, because in the scenario of the tetraquark states, the doubly-charmed tetraquark states with the isospin $(I,I_3)=(1,0)$ and $(0,0)$ have degenerated masses. Moreover, we make predictions for other scalar/axialvector/tensor  doubly-charmed tetraquark molecular states, which can be searched for in the $DD\pi $, $DD\gamma $, $DD\pi \pi$,  $DD_s\gamma$, $DD_s\pi\pi$,  $DD_s\pi\pi\pi$, $D_sD_s\gamma$,  $D_sD_s\pi\pi\pi$ invariant mass distributions  at the BESIII, LHCb, Belle II,  CEPC, FCC, ILC in the future.

\section*{Acknowledgements}
This  work is supported by National Natural Science Foundation, Grant Number  12175068.

\end{document}